\documentclass[3p,times, twocolumn]{elsarticle}
\usepackage{lineno,hyperref} \modulolinenumbers[5]

\journal{Physics Letters B}

\usepackage{graphicx} 
\usepackage{dcolumn}  
\usepackage{subfigure}
\usepackage{amsmath,amssymb}
\usepackage{color}
\usepackage{wasysym}

\newcommand{\R}    {}

\newcommand{\beq}{\begin{equation}}
\newcommand{\eeq}{\end{equation}}
\newcommand{\beqn}{\begin{eqnarray}}
\newcommand{\eeqn}{\end{eqnarray}}

\newcommand{\gevc}{\ensuremath{\,{\mathrm{GeV}}}}
\newcommand{\mevc}{\ensuremath{\,{\mathrm{MeV}}}}
\newcommand{\gev}{\ensuremath{\,{\mathrm{GeV}}}}
\newcommand{\mev}{\ensuremath{\,{\mathrm{MeV}}}}

\newcommand{\Br}{\ensuremath{\mathcal{B}}}

\newcommand{\pp}{\ensuremath{\psi^\prime}}

\newcommand{\ccbar}{\ensuremath{c\bar{c}}}

\newcommand{\pppi}{\ensuremath{\pp\pi^+}}
\newcommand{\kp}{\ensuremath{K^{\,-}\pi^+}}
\newcommand{\Km}{\ensuremath{K^{\,-}}}

\newcommand{\Ktw}{\ensuremath{K_{2}^{\,*}(1430)}}
\newcommand{\Kon}{\ensuremath{K^{\,*}(890)}}

\newcommand{\Dss}{\ensuremath{D_{s}^{\,*-}}}
\newcommand{\Ds}{\ensuremath{D_{s}^{\,-}}}

\newcommand{\Dsr}{\ensuremath{D_{s}^{\,\prime -}}}
\newcommand{\Dssr}{\ensuremath{D_{s}^{\,*\prime -}}}
\newcommand{\DsJol}{\ensuremath{D_{s1}^{\,*}(2700)^-}}
\newcommand{\Dssra}{\ensuremath{D_{s}^{\,(*)\prime -}}}

\newcommand{\Dsra}{\ensuremath{D_{s}^{\,(*)\prime -}}}

\newcommand{\Dst}{\ensuremath{D^{\,*}}}

\newcommand{\D}{\ensuremath{D}}

\newcommand{\Dstp}{\ensuremath{D^{\,*+}}}

\newcommand{\Dsta}{\ensuremath{\bar{\!D}{}^{\,*0}}}

\newcommand{\Dp}{\ensuremath{D^{\,+}}}

\newcommand{\Dn}{\ensuremath{D^{\,0}}}
\newcommand{\Da}{\ensuremath{\bar{\!D}{}^{\,0}}}

\newcommand{\DD}{\ensuremath{\Dsta\Dp}}
\newcommand{\DDp}{\ensuremath{\Dstp\Da}}
\newcommand{\DDa}{\ensuremath{(\bar{D}D\,)^{*+}}}

\newcommand{\aB}{\ensuremath{\bar{\!B}}}

\newcommand{\Z}{\ensuremath{Z^{\,+}}}
\newcommand{\Zp}{\ensuremath{Z(4430)^+}}

\newcommand{\tdec}{\ensuremath{\theta_{\mathrm {dec}}}}
\newcommand{\trot}{\ensuremath{\theta_{\mathrm {rot}}}}
\newcommand{\tform}{\ensuremath{\theta_{\mathrm {form}}}}

\begin{document}

\title{Charged charmonium-like $Z^{\,+}(4430)$ from rescattering in
  conventional $B$ decays}

\author{P.~Pakhlov}
\address{Moscow Institute of Physics and  Technology, Moscow Region, Russia}
\address{Moscow Physical Engineering Institute, Moscow, Russia}

\author{T.~Uglov}
\address{Moscow Physical Engineering Institute, Moscow, Russia}
\address{Moscow Institute of Physics and  Technology, Moscow Region, Russia}

\begin{abstract}
\noindent

In a previous paper we suggested an explanation for the peak
designated as \Zp\ in the \pppi\ mass spectrum, observed by Belle in
$\aB \to \pp \pi^+ K$ decays, as an effect of $\DD \to \pppi$
rescattering in the decays $\aB \to \Dsr \D$, where the \Dsr\ is an
as-yet unobserved radial excitation of the pseudoscalar ground state
\Ds-meson. In this paper, we demonstrate that this hypothesis provides
an explanation of the double $Z^+$-like peaking structures, which were
studied by LHCb with much higher statistics. While according to our
hypothesis, the origin of the peaking structures is due to the
kinematical reflection of conventional resonances in the unobserved
intermediate state, the amplitude of the \Zp\ peak carries a
Breit-Wigner-like complex phase, arising from the intermediate
\Dsr\ resonance. Thus, our hypothesis is entirely consistent with the
recent LHCb measurement of the resonant-like amplitude behavior of
the \Zp. We perform a toy fit to the LHCb data, which illustrates that
our approach is also consistent with all the observed structure in the
LHCb $M(\pppi)$ spectrum. We suggest a critical test of our hypothesis
that can be performed experimentally.
\end{abstract}


\maketitle

Many $XYZ$ states above open charm threshold, and decaying into
charmonium and light hadron(s) have been observed within the past
decade. Their conventional interpretation as charmonium states remain
controversial as their properties, especially their large decay rates
into final states without open charm, do not easily match the levels
of heretofore unobserved charmonia. Various exotic explanations, such
as tetraquarks, molecular states, charmonium hybrids and
hadrocharmonium are also not fully embraced by the physics community,
as they cannot describe the variety of observed states, and all their
measured properties, within a single self-consistent approach.

The first charmonium-like state, the \Zp, which is entirely
inconsistent with a simple charmonium interpretation, was observed by
Belle~\cite{belle1,belle2} in 2007 as a peak in the \pppi\ mass near
$M\sim4430\mevc$ in $B$ decays. Interpreted as a real resonance
containing a \ccbar\ pair, its minimal quark content given its
non-zero charge ($u\bar{d}\ccbar$), is necessarily exotic. The
existence of the \Zp\ was cast into doubt by BaBar~\cite{babar1}, but
the recent \Zp\ observation by LHCb~\cite{lhcb1} unambiguously (with
significance $\sim 14\sigma$) supports Belle's claim.

Among the exotic explanations of the \Zp, the most popular are the
tetraquark~\cite{tetraq}, hadrocharmonium~\cite{hadroc} and
$DD^{\,**}$ molecules~\cite{molec}. There are also non-resonant
interpretations such as the ``cusp effect''~\cite{cusp}, rescattering
via the chain $\aB \to D^{\,*-}D_1(2420) K \to \pppi K$~\cite{rosner},
and the initial single pion emission mechanism~\cite{ipe}. In our
previous paper~\cite{pakhlov}, we suggested another possible
explanation of the \Zp\ peak, resulting from $\DD \to \pppi$
rescattering in the decays $\aB \to \Dsr \Dp$.  Although this decay
has not yet been observed and even the \Dsr-meson not yet discovered,
the branching fraction for the decay $\aB \to \Dsr \Dp$ is expected to
be large, similar to that observed for $B^{\,+} \to \Dssr
D^{\,0}$~\cite{brodzicka}, while the mass of the \Dsr\ is predicted in
the range $(2600-2650)\mevc$\,---\,which corresponds to the range that
provides a \Zp\ peak value consistent with the extant experimental
data.

If our \emph{ad hoc} hypothesis is correct, the origin of the
\Zp\ peaking structure is caused by the presence of a conventional
resonance (the \Dsr\ meson) in the hidden intermediate state. However,
our explanation also implies an interesting underlying phenomenon:
namely, a non-vanishing rescattering amplitude over a wide range of
$M(\DD)$. In this Letter, we demonstrate that our approach is fully
consistent with all the experimental data, including the recent
\Zp\ phase study by LHCb, as the \Zp\ phase would then arise from the
Breit-Wigner \Dsr\ amplitude. We show that other structures that are
evident in the LHCb \pppi\ spectrum can be attributed to similar
effects. We also suggest here a critical test of our hypothesis that
can be performed by Belle, BaBar and LHCb.

First, we note that in our previous paper~\cite{pakhlov} we have
predicted the quantum numbers of the \Zp\ to be $J^P=1^+$ based on the
simple argument that the $\DD \to \pppi$ rescattering should be
dominated by $S$-waves in both the colliding \DD\ and also the
produced \pppi\ systems. This prediction was confirmed by subsequent
Belle~\cite{chilikin} and LHCb~\cite{lhcb1} measurements. We also
predicted the presence of other structures in the \pppi\ spectrum, in
particular near $M\sim 4200\mevc$, which arise from another $\aB \to
\Dssr D$ decay chain. Such a broad peak at $M = 4239$\mevc\ is,
indeed, observed in the LHCb data, and has been interpreted as another
\Z\ resonance.

\begin{figure}[thb]
\begin{center}
\includegraphics[width=0.46\textwidth] {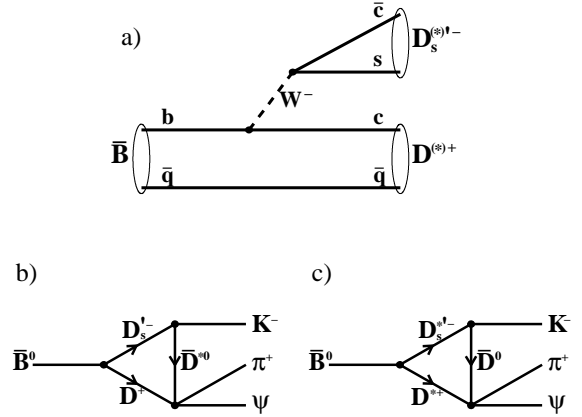} \\
\end{center}
\caption{a) Feynman diagram for $B$ decay into radially excited
  \Ds\ mesons. Rescattering processes $\DDa \to \psi \pi^+$,
  represented by triangle diagrams: a) for the chain~\ref{react1} and
  b) for the chain~\ref{react2}. }
\label{f_diag}
\end{figure}

We reiterate the main points of our hypothesis. As in our previous
paper~\cite{pakhlov} we consider $B$ decays governed by the tree
diagrams shown in Fig.~\ref{f_diag}\,a). In these decays the $W^-$ is
directly coupled to the radial excitations of the \Ds\ and \Dss-mesons
(the \Dsr\ and \Dssr) in a similar way as to their ground states. One
of such mode, $B^{\,-} \to \DsJol \Dn$, was observed by Belle with a
relatively large branching fraction $\Br(B^{\,-} \to \DsJol \Dn)
\times \Br(\DsJol \to \Da K) \sim 10^{-3}$~\cite{brodzicka}. The
measured quantum numbers of the \DsJol\ ($J^P=1^{-}$) suggest the
interpretation of this state as the \Dssr\ meson. Other channels and
even the \Dsr\ have not, thus far, been explicitly searched for
experimentally. However, the inclusive $B \to \D^{(*)} \bar{D}{}^{(*)}
K$ branching fractions are large: they vary from 0.1 to
1\%~\cite{pdg}. It is natural to assume that they should be saturated
by two-body modes with intermediate radial \Ds\ and \Dss\ excitations,
since the known contribution of orbital \Ds\ excitations to these
final states is small~\cite{pdg}.

The \Dsr-meson is expected to decay mostly to the $\Dst K$ final
state, as the decay $\Dsr \to D K$ is forbidden by parity
conservation, while the \Dssr\ decays to both $D K$ and $\Dst
K$~\cite{pdg}. Therefore the $B$ decays under consideration hadronize
into $D^{(*)} \bar{D}{}^{(*)} K$ final states. We note that two
charmed mesons are produced spatially at the same point and fly apart
relatively slowly with $v/c \approx 0.3-0.5$. Therefore one can expect
the non vanishing rescattering of two charmed mesons into charmonium
plus a light meson. Considering the $S$-wave rescattering as a
recombination of the charm quark from one charmed meson and the charm
antiquark from the other into charmonium, with the simultaneous
merging of a light quark-antiquark pair into a light meson, we
conclude that only $D\bar{D}^*+c.c.$ states can result in rescattering
into \pppi\ or $J/\psi \pi^+$. Other ($D\bar{D}$ and $D^*
\bar{D}{}^*$) can rescatter to other charmonia and/or other light
mesons. The rescattering amplitude can be determined by the overlap
integral of two products of wave functions with the same quark
content, taking into account color suppression. We do not attempt such
calculations, which can only be done by invoking a model for light and
heavy mesons and charmonium wave functions, but simply assume that
this amplitude is small but not vanishing, and does not change
dramatically within the range of interest ($M_\D + M_{\Dst} < M_{\DDa}
\lesssim 4.8\gevc$).

\begin{figure*}[bht]
\begin{center}
\includegraphics[width=0.76\textwidth] {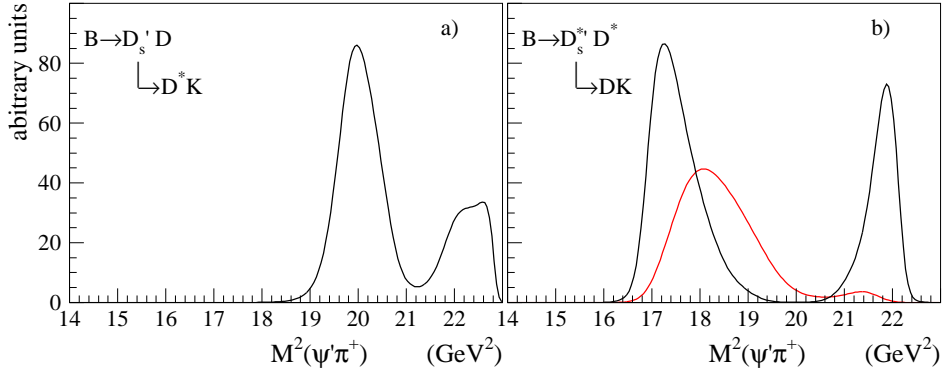}
\end{center}
\caption{The $M_{\pppi}^{\,2}$ spectrum in the decay $B \to \Dssra
  \Dp$, followed by rescattering $\DDa \to \pppi$, calculated
  according to Equation~(\ref{amp}). a) and b) correspond to the
  chains (\ref{react1}) and (\ref{react2}). The black and red curves
  correspond to the lineshapes for $\lambda_{\Dst}=0$ and
  $\lambda_{\Dst}=\pm1$, respectively.}
\label{fig1}
\end{figure*}

Of the decay chains discussed above, only two can contribute to the
$\pppi K$ final state:
\begin{equation}
\aB \to \Dsr \Dp, ~ \mathrm{followed~by} ~ \Dsr \to \Dsta \Km \, ,
\label{react1} 
\end{equation}
and 
\begin{equation}
\aB \to \Dssr \Dstp, ~ \mathrm{followed~by} ~ \Dssr \to \Da \Km \, .
\label{react2}
\end{equation}
They corresponds to the triangle diagrams in Fig.~\ref{f_diag}\,b) and
c) respectively. The decay $\aB \to \Dssr \Dp$ which could otherwise
contribute to this process, has parity opposite $\Z \Km$ and is
therefore not considered. We note that while parity is not conserved
in $B$ decays, the rescattering process is mediated by the strong
interaction and requires parity conservation.

We introduce a common notation, \DDa, to refer to both \DD\ and
\DDp\ systems in the reactions (\ref{react1}) and (\ref{react2}),
respectively, and designate as \Z\ a pseudoparticle with $J^P=1^+$
formed by the \DDa\ combination before its subsequent decay to \pppi.

As in our previous paper, we calculate the amplitude of interest in
the on-shell approximation of the triangle diagrams
(Fig.~\ref{f_diag}), taking into account the \Dssra\ Breit-Wigner
amplitude. We also include the \Dst\ spin rotation amplitudes, which
provide the proper \Dst\ helicity in the \Z\ system, corresponding to
$S$-wave formation of the \Z.
{\R Depending on the \Dsra\ decay angle
different values of \Dsra\ mass within the Breit-Wigner distribution
can yield the same \DDa\ mass. Thus, the total amplitude
$\mathcal{A}_{\Z}$ should be calculated as a superposition of all
allowed values of $M(\Dsra)$, accounting for the variation in phase
with mass. Unlike our previous paper, here we therefore integrate
the entire allowed kinematic region, explicitly including the
variation in phase.}
 This procedure is more
rigorous, and the \Z\ shape is also slightly changed, relative to our
previous calculations. The full decay amplitude has the following form
in the helicity formalism:
\begin{eqnarray}
\begin{aligned}
\mathcal{A} & (M_{\Z}  \! \equiv \, M_{\DDa})\!   =  \!  \sum_\lambda \! \int \!  
\mathcal{A}_{BW} (M_{\Dsra})  \\
& D^{J}_{0,\lambda} (\tdec) ~ D^1_{\lambda,0}(\trot) ~ D^1_{0,0} 
(\tform) ~ d M_{\Dsra} \,, 
\end{aligned}
\label{amp}
\end{eqnarray}
where $J$ is the \Dssra\ spin; \tdec\ is the decay angle of the
\Dssra\ (the angle between the \aB\ and $\bar{D}{}^{(*)0}$ in the
\Dssra\ rest frame); \trot\ is the rotation angle of the \Dsta\ spin
from the \Dssra\ frame for the reaction~(\ref{react1}) or the
\aB\ frame for the reaction~(\ref{react2}) to the \Z\ frame;
\tform\ is the formation angle of \Z, \emph{i.e.} the angle between
the \aB\ and \Dst\ in the \Z\ rest frame.  The first Wigner
$D$-function is responsible for the proper angular distribution of the
\Dssra\ decay (in the case considered in (\ref{react2}), only the zero
helicity projection is considered). The second function,
$D^1_{0,\lambda}(\trot)$, describes the \Dst\ spin rotation from the
frame where it is produced to the frame where it is absorbed. Finally,
the $D^1_{0,0} (\tform)$ corresponds to the proper formation of the
spin-1 \Z\ pseudostate from the vector (\Dst) and the pseudoscalar
(\D). Two variables, $M_{\Dsra}$ and \tdec, fully describe the
three-body kinematics, thus $M_{\DDa}$, \trot\ and \tform\ are
functions of these two variables.

We have performed the calculation of Equation~(\ref{amp}) numerically
using Monte Carlo simulations. We first generate the $\aB \to \Dsra
\D^{\,(*)}$ decay kinematics. The mass and width of the \Dssr\ are
fixed to the PDG values ($M=2.709\mevc$,
$\Gamma=0.112\mev$~\cite{pdg}); the \Dsr\ parameters are fixed to
$M=2610\mevc$ and $\Gamma=100\mev$ as in our previous
paper~\cite{pakhlov} (the expected $2S^1 - 2S^3$ splitting is
$(60-100) \mevc$~\cite{theor}). For each generated event, we then
calculate the expected contribution to the full amplitude according to
Equation~(\ref{amp}) (this amplitude is a function of the kinematic
characteristics of a particular event). Finally, we sum over (complex)
amplitudes corresponding to the same $M_{\DDa}$ bin. The resulting
\Z\ shapes (equal to $|\sum \mathcal{A}|^2$) for the chains
(\ref{react1}) and (\ref{react2}) are shown in Fig.~\ref{fig1}; for
the latter we plot separately the contributions of different
\Dst\ helicities.

\begin{figure*}[bth!]
\begin{center}
\includegraphics[width=0.76\textwidth] {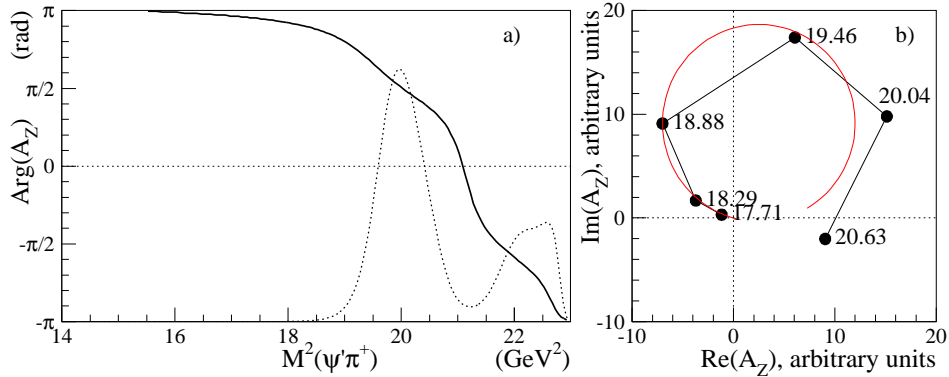} 
\end{center}
\caption{ a) The phase of $\mathcal{A} (M_{\Z})$ in the decay $B \to
  \Dsr (\to \Dsta \Km) \Dp$, followed by rescattering $\Dsta\Dp \to
  \pppi$, calculated according to Equation~(\ref{amp}) {\R as a
    function of $M_{\Z}$; the dashed curve represents the process
    lineshape ($\propto |\mathcal{A} (M_{\Z})|^2$).  b) The Argand diagram
    for rescattering contribution around the \Zp\ peak.} }
\label{fig2}
\end{figure*}

The phase of the \Z\ amplitude, $\mathrm {arg} (\mathcal{A}_{\Z})$,
from the reaction~(\ref{react1}), which is responsible for the most
prominent peak of the \Zp, is presented in Fig.~\ref{fig2}\,a).  {\R
  Equivalently, we plot the Argand diagram Fig.~\ref{fig2}\,b) using
  the same $M_{\Z}$ binning as the LHCb experiment for direct
  comparison. The initial phase in our case is arbitrarily set to
  $\pi$, while for the LHCb experiment, it is fixed relative to the
  reference $B\to \pp K$ phase from their $4D$-fit. The phase
  variation around the \Zp\ peak arises from the \Dsra\ Breit-Wigner
  phase variation via the convolution with the angular variables in
  Equation~(\ref{amp}). We note that the higher mass region of
  \Dsra\ corresponds to lower \Zp\ mass, and vice versa. Therefore, in
  the region} 
around the \Zp\ the phase turns out to have opposite behavior relative
to the conventional Breit-Wigner definition: it tends to rotate
clockwise in the Argand diagram. However, experimentally the direction
of amplitude rotation cannot be determined as there is a two-fold
ambiguity ($\mathcal{A} \leftrightarrow \bar{\mathcal{A}}$) in the
extraction of the \Z\ amplitude from the measured $\left|
\mathcal{A}_{\Z} + \mathcal{A}_{{\mathrm{non-}}\Z} \right|^2$. Thus,
our hypothesis is fully consistent with the LHCb Argand diagram.

To further illustrate that our hypothesis is plausible, we use the LHCb
\pppi\ mass spectrum with vetoed \Kon\ and \Ktw\ resonances (Fig.~4
from~\cite{lhcb1}) and perform a toy fit to this spectrum ignoring
interference between major $B \to \pp K^{\,*(*)}$ and rescattering
contributions.
\begin{figure}[htb]
\begin{center}
\includegraphics[width=0.46\textwidth] {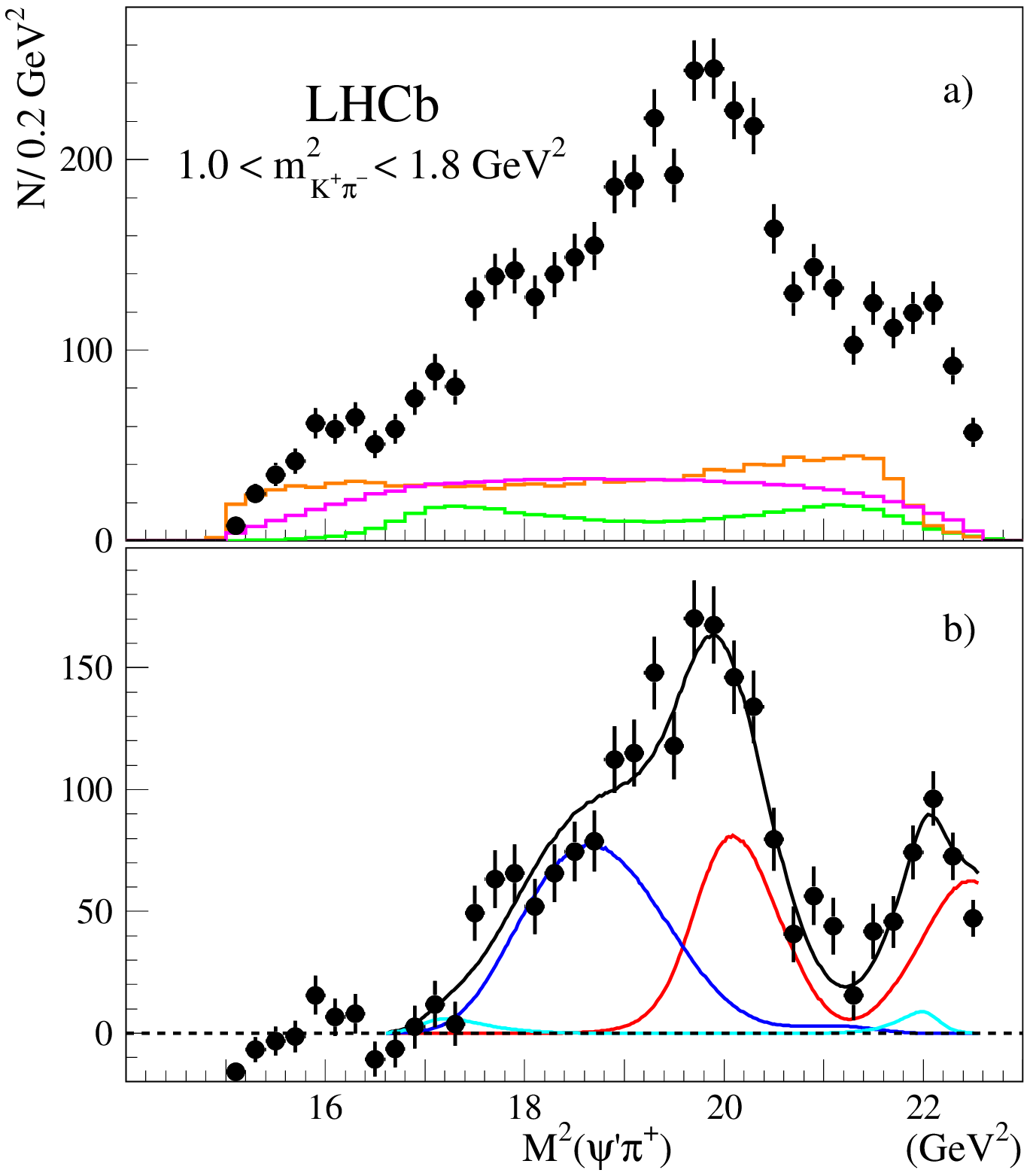} 
\end{center}
\caption{a) Distribution of $M_{\pppi}^{\,2}$ in the LHCb data for
  $1.0 < M^{\,2}_{\kp} <1.8\gev^2$ borrowed from~\cite{lhcb1} (black
  points); the orange, green and magenta histograms are contributions
  from \Kon, \Ktw\ and $S$-wave three-body phase space, respectively,
  expected from the LHCb fit. b) Distribution of $M_{\pppi}^{\,2}$
  after incoherent subtraction of contributions from \Kon, \Ktw\ and
  non-resonant three-body decays. The black curve represents our fit
  to the data points. The red, blue and cyan curves represent
  contributions from the (\ref{react1}) and (\ref{react2}) processes,
  with $\lambda=1$ and $\lambda=0$, respectively.}
\label{lhcb_m}
\end{figure}
This is not a fully correct procedure, we thus use it for illustration
only, but having access to the published one-dimensional
$M_{\pppi}^{\,2}$ projections only, we can not calculate
phase-dependent interference effects. We first estimate the remaining
contributions from \Kon, \Ktw\ and $S$-wave three-body phase space,
after selecting the $1.0 < M^{\,2}_{\kp} <1.8\gev^2$ interval, using
Figs.~3\,a) and b) from~\cite{lhcb1}. The LHCb data points with these
three contributions superimposed (the histogram colors correspond to
the LHCb notation) are shown in Fig.~\ref{lhcb_m}\,a). The spectrum in
Fig.~\ref{lhcb_m}\,b) is obtained after a bin-by-bin subtraction of
$K^{\,*(*)}$ and non-resonant three-body decays. We attribute the
remaining spectrum to the rescattering contribution and perform a fit
to this spectrum with a sum of contributions from the reactions
(\ref{react1}) and (\ref{react2}), therefore with five free
parameters. We note that all intermediate $B$ decay channels with
various \Dssra\ states contribute to \Z\ production coherently with
the same universal rescattering amplitude. The fit results are plotted
in Fig.~\ref{lhcb_m}\,b) with the black solid line, and nicely
describe all the features observed in data.

\begin{figure}[htb]
\begin{center}
\includegraphics[width=0.46\textwidth] {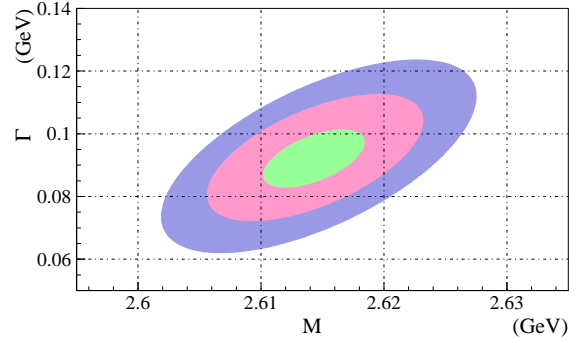}
\end{center}
\caption{Mass {\it vs.} total width of the \Dsr\ resonance predicted
  from its contribution to the rescattering diagram.}
\label{m-wid}
\end{figure}

We estimate the parameters of the \Dsr\ meson from the fit to the LHCb
data. We vary the \Dsr\ mass and width and calculate the confidence
level of the fit for each set of values. The result of this exercise
is presented in Fig.~\ref{m-wid}, where the green, magenta and blue
contours correspond to $1\sigma$, $2\sigma$ and $3\sigma$ levels,
respectively. The \Dsr\ parameters turn out to be well statistically
constrained by the fit: $M=(2614\pm 4)\mev$,
$\Gamma=(92\pm10)\mev$. However, there is a systematic uncertainty in
these values due to the effect of interference with the $K^{*(*)}$
background. To estimate this effect we ascribe different phases to the
amplitudes of \Kon, \Ktw\ and $S$-wave three-body phase space and
perform another fit to the distribution in Fig.~\ref{lhcb_m}\,a) with
varying \Dsr\ mass and width. Variations of the best fit
\Dsr\ parameters depending on the $K^{*(*)}$ phases are estimated to
be $\pm 10\mev $ for the \Dsr mass and $^{+20}_{-13}\mev$ for its
width. We thus conclude that, to explain the \Zp\ peak, the parameters
of the \Dsr\ meson should be in the interval: $M=(2614\pm 4
^{+20}_{-13})\mev$, $\Gamma=(92\pm10 \pm 10)\mev$.

Soon after this paper was submitted, another experimental analysis of
$\aB \to J/\psi \kp$ by Belle appeared~\cite{chilikin2}. The existence
of the broad structure at $M(J/\psi \pi^+)\sim 4200\mev$ is
established in that measurement with high significance and with
preferred assignment of the quantum numbers $J^P = 1^+$; strong
evidence for a \Zp\ signal is also found. The parameters of the two
bumps are consistent between the $J/\psi \pi^+$ and
\pppi\ analyzes. However, their relative phases with respect to $B \to
\psi K^{*(*)}$ background look different, ({\it e.g.} the \Zp\ peak is
seen as destructively interfering). While in our approach only the
strengths of the \DDa\ rescattering amplitudes, which are real
numbers, into $J/\psi \pi$ and \pppi\ can be different, this fact can
be attributed to the different phases of the interfering $K^{(*(*))}$
background amplitude under \Z's in these two modes. Indeed, the 3-body
phase space is different due to the different $J/\psi$ and
\pp\ masses. Thus, not only the different helicity regions of the same
$K^{(*(*))}$ contribute to the \Z\ regions in these two modes, but
also relative contributions of allowed $K^{**}$ may differ.

A real test of our hypothesis can be achieved with a 4D-fit performed
by Belle, BaBar and LHCb for $B\to \pppi \Km$ decays using
amplitudes~(\ref{amp}) instead of resonance-like \Z's.  Obviously the
fitting model with rescattering includes many free parameters: at
least three complex amplitudes to describe all possible contributions
as well as the as-yet-undetermined parameters of the \Dsr\ resonance.
It is important to fix these amplitudes using a study of $B \to \Dsta
\Dp \Km$ and $B \to \Da \Dstp \Km$, which is possible at B-factories
or LHCb. However, there is an easier way to check our hypothesis
experimentally. The \Z-like structures should appear in the
distributions of $M_{(\Dst_\perp \bar{D})^+}\times \cos^2(\tform)$ in
either $B \to \Dsta \Dp \Km$ or $B \to \Da \Dstp \Km$ decays, or in
both. The $M_{(\Dst_\perp \bar{D})^+}\times \cos^2(\tform)$ is the
$(\Dst_\perp \bar{D})^+$ combination mass spectrum corrected in each
bin for the fraction of the $\Dst$ transverse component in the
\DDa\ rest frame, and also the $1^+$ formation factor
$D^2(\tform)=\cos^2(\tform)$.

In summary, we show that $\DD \to \pppi$ rescattering in the decay
chain $\aB \to \Dsr \Dp$, $\Dsr \to \Dsta \Km$ can explain the
appearance of an observed peak in the \pppi\ mass spectrum in $\aB \to \pppi
\Km$ decays around $M \sim 4430\mevc$ and also correctly describes the
quantum numbers and amplitude resonance-like behavior. This approach
allows also to describe another peak at $M\sim 4.2\gevc$ 
observed in LHCb data and which has been interpreted as another exotic
resonance, as well as a high mass structure at the upper bound of the
mass spectrum, which remains still undersaturated by the LHCb fit
(with many $K^{\,**}$ and two \Zp's included).
 
The authors thank Yu. Kalashnikova for useful comments.
This work is supported in part by RFBR grant 15-02-99495.

\end{document}